\documentclass{article}

\usepackage{arxiv}

\usepackage[utf8]{inputenc} 
\usepackage[T1]{fontenc}    
\usepackage{hyperref}       
\usepackage{url}            
\usepackage{booktabs}       
\usepackage{amsfonts}       
\usepackage{nicefrac}       
\usepackage{microtype}      

\usepackage{cleveref}       
\usepackage{lipsum}         
\usepackage{graphicx}
\usepackage{natbib}
\usepackage{doi}
\usepackage{wrapfig}
\usepackage{xcolor}

\title{Human-AI Teaming Using Large Language Models: \\Boosting Brain-Computer Interfacing (BCI) and \\ Brain Research}

\newif\ifuniqueAffiliation
\uniqueAffiliationtrue

\ifuniqueAffiliation 
\author{ \href{https://orcid.org/0000-0001-7206-9304}{\includegraphics[scale=0.06]{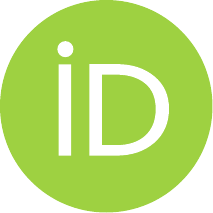}\hspace{1mm}Maryna Kapitonova}\thanks{www.neuromentum.ai} \\
	Department of Neurosurgery\\
    University Hospital Freiburg, Germany\\
	NeuroMentum AI\\
    \texttt{kapitonova@neuromentum.ai} \\
	\And
	\href{https://orcid.org/0000-0002-4993-466X}{\includegraphics[scale=0.06]{orcid.pdf}\hspace{1mm}Tonio Ball} \\
	Department of Neurosurgery\\
    University Hospital Freiburg, Germany\\
	NeuroMentum AI\\
	\texttt{ball@neuromentum.ai} \\
}
\else
\usepackage{authblk}

\newbox{\orcid}\sbox{\orcid}{\includegraphics[scale=0.06]{orcid.pdf}} 
\author[1]{%
	\href{https://orcid.org/0000-0000-0000-0000}{\usebox{\orcid}\hspace{1mm}David S.~Hippocampus\thanks{\texttt{hippo@cs.cranberry-lemon.edu}}}%
}
\author[1,2]{%
	\href{https://orcid.org/0000-0000-0000-0000}{\usebox{\orcid}\hspace{1mm}Elias D.~Striatum\thanks{\texttt{stariate@ee.mount-sheikh.edu}}}%
}
\affil[1]{Department of Computer Science, Cranberry-Lemon University, Pittsburgh, PA 15213}
\affil[2]{Department of Electrical Engineering, Mount-Sheikh University, Santa Narimana, Levand}
\fi

\renewcommand{\headeright}{Preprint}
\renewcommand{\undertitle}{Preprint}
\renewcommand{\shorttitle}{Human-AI Teaming for BCI and Brain Research}

\hypersetup{
pdftitle={A template for the arxiv style},
pdfsubject={q-bio.NC, q-bio.QM},
pdfauthor={David S.~Hippocampus, Elias D.~Striatum},
pdfkeywords={First keyword, Second keyword, More},
}

\begin{document}
\maketitle

\begin{abstract}
	Recently, there is an increasing interest in using artificial intelligence (AI) to automate aspects of the research process, or even autonomously conduct the full research cycle from idea generation, over data analysis, to composing and evaluation of scientific manuscripts. Examples of working \textit{“AI scientist”} systems have been demonstrated for computer science tasks and running molecular biology labs. While some approaches aim for full autonomy of the scientific AI, others rather aim for leveraging human-AI teaming. Here, we address how to adapt such approaches for boosting Brain-Computer Interface (BCI) development, as well as brain research resp. neuroscience at large. We argue that at this time, a strong emphasis on human-AI teaming, in contrast to fully autonomous “AI BCI researcher,” will be the most promising way forward. We introduce the collaborative workspaces concept for human-AI teaming based on a set of \textit{Janusian} design principles, looking both ways, to the human as well as to the AI side. Based on these principles, we present ChatBCI, a Python-based toolbox for enabling human-AI collaboration based on interaction with Large Language Models (LLMs), designed for BCI research and development projects. We show how ChatBCI was successfully used in a concrete BCI project on advancing motor imagery decoding from EEG signals. Our approach can be straightforwardly extended to broad neurotechnological and neuroscientific topics, and may by design facilitate human expert knowledge transfer to scientific AI systems in general.

\end{abstract}


\section{Introduction}

Human–Artificial Intelligence (AI) collaborations may play an important role in promoting scientific discovery \cite{wang2023scientific}. Efforts to automate research and development tasks have recently led to the creation of \textit{“AI scientists”—AI agents specialized to autonomously carry out various roles in the scientific process.} AI scientists may run virtual or even physical labs (self-driving labs, SDLs \cite{rapp_self-driving_2024}) and thus help to \textit{“super-charge”} fields including biomedical research, material science, or computer science.

For example, in an effort to leverage AI for biomedical research, Swanson and colleagues have recently created a virtual laboratory that brings together multiple AI scientists—implemented as large language models (LLMs) with specialized roles—to collaborate autonomously on objectives defined by human investigators. This system rapidly proposed nearly 100 designs for antibody fragments, known as nanobodies, capable of binding to the virus responsible for COVID-19—accomplishing in a short span what might otherwise take a purely human team much longer \cite{swanson_virtual_2024}. In the field of computer science, Lu and colleagues introduced an AI scientist system that aims to automate the complete research cycle. They also use a LLM to survey scientific literature, formulate hypotheses, test them by running experiments, and then compile the results into a written paper. The system goes a step further by performing its own \textit{‘peer review,’} evaluating and refining the outputs it generates \cite{lu2024ai_scientist}. However, despite these impressive first demonstrations, substantial challenges remain for AI-driven scientific discovery in general, as well as for the adaptation of AI-scientists to the unique requirements that the scientific domains of research and development may impose on such a system.

For AI-scientist systems (AISS) in general, for example, one major challenge is how to integrate the vast amount of implicit, unstructured, and often context-dependent knowledge and expertise that human scientists rely on for experimental design, interpretation, and innovation—which may not be sufficiently represented in the available training data. For example, Rihm and colleagues argue that current AI-driven labs are only able to incorporate explicit knowledge such as recorded data, scientific documentation, and publications and hence only see the \textit{“tip of the iceberg”} of the relevant knowledge \cite{rihm_transforming_2024}.

\begin{wrapfigure}{l}{0.5\textwidth} 
    \centering
    \includegraphics[width=0.48\textwidth]{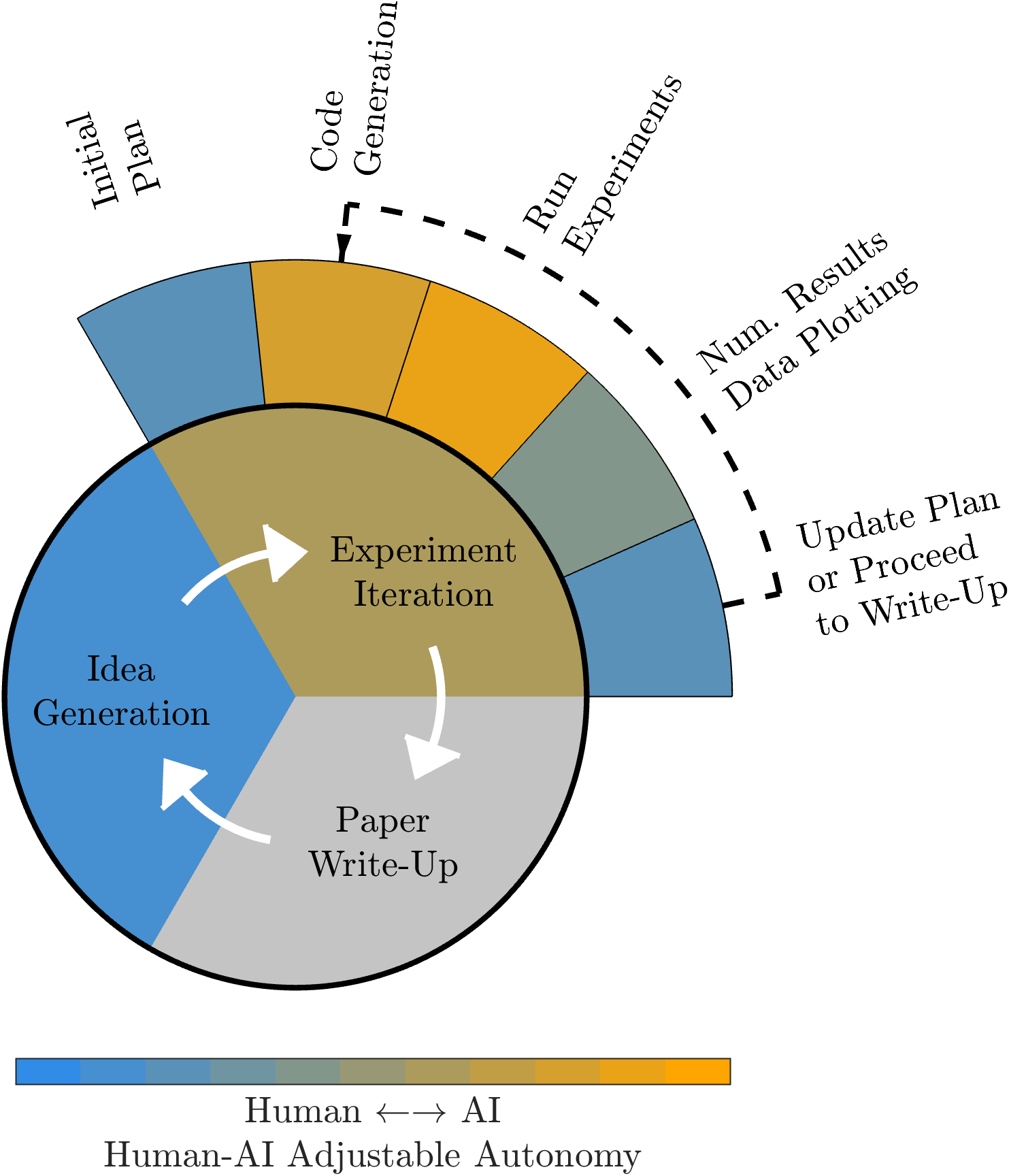} 
    \caption{A conceptual illustration of AI-driven scientific research processes for Brain-Computer Interfacing (BCI) and brain research, based on the concept as proposed by \cite{lu2024ai_scientist}. In contrast to their concept, we consider numerical results and data visualization as an integral part of the experimental iteration loop. Also in contrast to their fully autonomous approach, we support flexible levels of shared autonomy for each phase of the cycle (color coded to reflect the example BCI project as reported here; automatic paper write-up was not addressed)}
    \label{fig:intro_ai_research}
\end{wrapfigure}

However, even explicit knowledge such as that expressed in scientific publications may hold problems of its own when used as training data for scientific AI systems. Geoffrey Boulton and Moumita Koley observed in a contribution for the International Science Council: \textit{“Many scientific journals and papers maintain high standards, too many lack proper editorial oversight, many lack rigour and integrity, some engage in fraudulent practices, [...], and agreed standards for overall governance of the process are lacking.”} \cite{council_science_blog}. All of these point might compromise training data quality. This situation, together with the recent \textit{“publishing explosion”} (47\% growth between 2016 and 2022 in the global number of published papers \cite{hanson_strain_2024}; nearly 250\% increase in the number of Google Scholar hits for the term \textit{“Brain-Computer Interface”} in the same time interval) which may reflect a overemphasis of quantity compared to quality, may result in quality issues with respect to the available training data for scientific AI systems. Such problems may be more or less serious in some research fields compared to others \cite{bakker_rules_2012}.

Thus, in addition to general challenges such as those related to training data availability and quality, adaptation of scientific AI to a specific research field may hold its own challenges—not only due to differences in research standards. AI systems may face vastly different requirements depending, for example, on the role of experimental work, the complexity of the relevant mathematical tools and concepts, or the degree of adoption of open science practices with resulting differences in public data and scientific software access. For example, adoption of scientific AI agents in subfields of machine learning such as diffusion modeling, transformer-based language modeling, and learning dynamics \cite{lu2024ai_scientist} benefits from publicly accessible and well-documented datasets of, for many relevant questions, adequate size and quality, as well as from well-established benchmarks. In contrast, fields such as EEG-based BCI research face much larger problems of small and diverse datasets of variable data quality. Additional domain-specific challenges arise due to the complexities of the foundation of EEG signals in multiscale brain dynamics and their relation to human cognition, emotion, and behavior \cite{friston_theory_2005}.

Against this background, our objective in the present work is to adopt AI scientific agents to the field of human brain research and Brain-Computer Interfacing (BCI) Figure~\ref{fig:intro_ai_research}. To address the general as well as the domain-specific challenges of this task, we introduce a concept for human-AI workspaces designed to foster scientific human-AI co-learning and collaboration, summarized in our \textit{Janusian Design Principles.} As an instance of this design concept, we present the LLM-based ChatBCI toolbox.

We show how ChatBCI enabled a human-AI team to perform a prototypical scientific BCI project based on a public EEG dataset in a very time-efficient and productive manner, addressing the full cycle of project phases from data import, exploration, and validation, over machine learning model selection and implementation (a deep convolutional neural network as the LLM-generated brain signal decoder), setting up a training loop with integrated data augmentation, to the joint interpretation of the obtained results. We illustrate how our structured interaction framework flexibly supports different levels of agent autonomy and demonstrate how it allows transfer of expert knowledge about non-obvious, non-trivial aspects of the EEG dataset used, from the human expert to the ChatBCI AI agent. This, in turn, proved a valuable teaching and training partner for the human side. Thus, we achieved meaningful human-AI co-learning together with substantially faster collaborative completion of our BCI research task than would have been possible with a purely human approach.
Our approach holds great potential for teaching and training in the field of human brain research and BCI development. LLM-generated brain signal decoders also present a novel class of AutoML for brain signal analysis. The \textit{“AI brain researcher”} or \textit{“AI BCI researcher”} presented here is an instance of an \textit{“AI worker”} in general. It will be interesting to study how abilities learned to handle the complexities of the human brain, and of the signals that can be measured from it, may also be helpful for AI workers in other domains. We anticipate that collaborative human-AI research and development in the field of brain research and BCI, as enabled by our design concept and ChatBCI toolbox, will pave the way toward AI systems with a deep and holistic understanding of human brain function - a \textit{"brain-grokking AI"} - with a far-reaching application potential.

\begin{wrapfigure}{l}{0.5\textwidth} 
    \centering
    \includegraphics[width=0.4\textwidth]{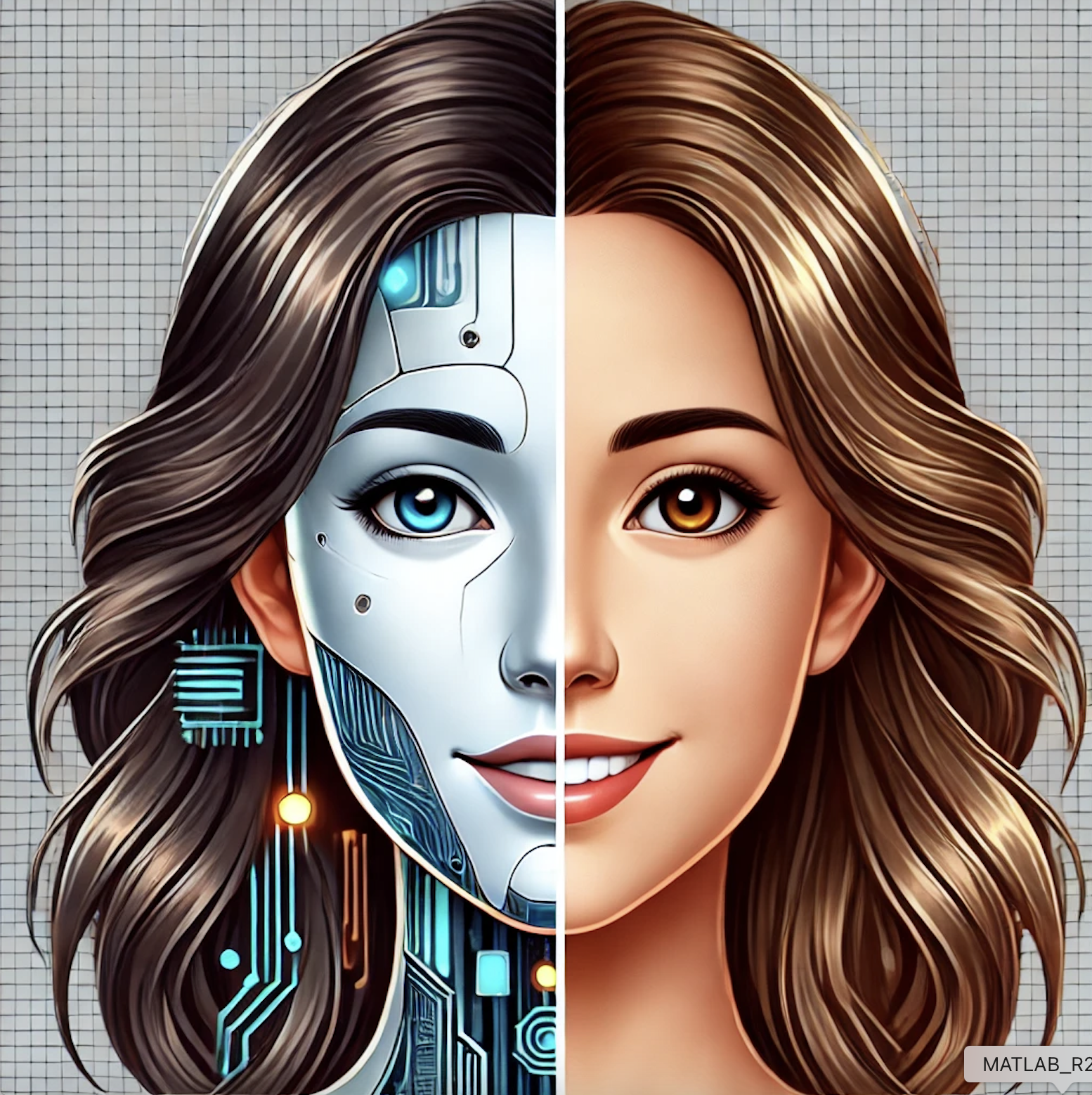}
    \caption{In the\textbf{ Janusian Vision} in designing human-AI workspaces, we embrace a dual-facing approach: one face directed toward \textbf{empowering human expertise}, and the other toward \textbf{amplifying AI capabilities} (DALL-E rendering of this idea).}
    \label{fig:Janusian}
\end{wrapfigure}

\section{ChatBCI toolbox overview and Janusian Design Principles}
\label{chatbci_overview}

The ChatBCI toolbox\footnote{The toolbox and corresponding detailed documentation will be made publicly available.} is fully implemented in Python, deep learning functionality is relying on PyTorch; in the present work we use GPT-4o as underlying LLM \cite{openai2024gpt4o}. It comprises the following main components: 
\begin{itemize}
    \item A collection of public EEG data sets in ChatBCI compatible format, including documented data validation
    \item EEG and BCI knowledge base
    \item Core functionality: EEG preprocessing, analysis, decoding models, training, interpretability, and visualization
    \item Auxiliary components and supporting infrastructure, such as for generating summaries of data and code directory structures and contents at adjustable levels of granularity. 
    \item LLM communication tools.
\end{itemize}

ChatBCI is by-design build to fostering human-AI collaboration (Figure~\ref{fig:Janusian}). To this aim, we have defined a set of design principles which are the foundation of our general framework for human-AI teams. In the following we give a summary of our current working version of these principles. \footnote{Named after Janus, two-faced Roman God of duality, to express the basic aspect of our design concept of looking both ways - to the human side, and to the AI side. We see this duality not as a contradiction but as a synthesis}.

\clearpage

\subsection*{Working version of the seven Principles of Janusian Design}

\begin{enumerate}
    \item \textbf{Speaking the Same Language:}  
    Janusian workspaces foster the development of a joint collaboration language, minimizing misunderstanding and confusion. This enables AI systems to act as intuitive partners, aligning with human intentions and promoting seamless interactions. To achieve this, the workspace leverages intuitive interfaces, contextual feedback, and iterative refinement, ensuring that communication between humans and AI becomes progressively more natural and effective.
    
    \item \textbf{Transparency and Trust:}  
    Trust requires mutual clarity and predictability. Explainability is a cornerstone, allowing users to progressively build confidence in AI systems by understanding their decision-making processes. The framework incorporates \textbf{bidirectional transparency}, encouraging not only explainable AI (XAI) methods, but also mechanisms for AI systems to ask clarifying questions when encountering ambiguous inputs or tasks. This fosters a collaborative dynamic where both human and AI agents seek to represent/understand each other’s intentions, enhancing mutual trust and reducing the likelihood of errors.
    
    \item \textbf{Building a Shared Knowledge Base:}  
    The workspace amplifies human and AI collaboration by creating and maintaining a shared knowledge base. This repository encodes specialized expert knowledge and evolves iteratively. The \textbf{Janusian Wiki (Jiki)} serves as the foundation for this knowledge base, providing a structured, centralized system for documenting workflows, best practices, and domain-specific knowledge. By linking tasks, tools, and insights, the Jiki ensures that both human collaborators and AI systems have access to the collective intelligence of the workspace, facilitating informed decision-making and continuous improvement.
    
    \item \textbf{Joint Integration of Priorities:}  
    The Janusian framework acknowledges that human-AI teams must balance and integrate priorities jointly, addressing trade-offs such as the need for immediate actions versus long-term objectives, or reconciling priorities driven by strategic considerations or overarching research questions ("top-down") with reactive, task-specific responses ("bottom-up"). 
    
    \item \textbf{Adaptive Autonomy:}  
    AI systems adjust their level of autonomy based on task demands, enabling granular human supervision where needed while offering higher degrees of independence for routine or low-stakes tasks. For example, in the ChatBCI toolbox, the adaptive autonomy principle is exemplified through \textbf{Collaborative Test-Driven Development (TDD)}, where the AI autonomously generates test cases and mock data for routine validations, while human oversight is applied to critical or ambiguous scenarios. This flexibility ensures that the level of autonomy dynamically aligns with the complexity and importance of the task.
    
    \item \textbf{Accessibility from Novice to Expert:}  
    The workspace aims to balance accessibility for novice users, such as those in teaching and training scenarios, with maintaining advanced functionality tailored for expert users. It also differentiates how information from these scenarios is processed and utilized on the AI side, ensuring that insights from both novice interactions and expert workflows are effectively integrated. This inclusive design promotes efficiency and reduces the supervision workload, enabling seamless collaboration across varying levels of expertise.

    \item \textbf{Continuous Evolution:}  
    A shared Janusian workspace is not static, but designed to foster mutual co-learning between humans and AI. This principle ensures that the ChatBCI toolbox or other workspaces based by-design on this concept not only supports effective and scalable data analysis but also fosters a dynamic workspace where both human and AI partners can co-evolve to achieve aligned goals.
\end{enumerate}

\noindent In summary, the Janusian framework establishes a synergistic foundation for human-AI collaboration by aligning intuitive interfaces, transparent processes, and adaptive autonomy with evolving user needs. In the following, we illustrate how the ChatBCI toolbox based on these design ideas can be used in a concrete project example. 

\section{Analyzing the BCI competition IV2a dataset using ChatBCI}
\label{bci_analysis}

As illustrated in Figure~\ref{fig:intro_ai_research}, we propose a conceptual framework for AI-driven scientific research processes tailored to Brain-Computer Interfacing (BCI) and brain research. This concept builds upon the approach introduced by \cite{lu2024ai_scientist}, which emphasizes fully autonomous AI systems executing all phases of the research cycle. In contrast, our approach highlights the importance of integrating numerical results and data visualization as essential elements within the experimental iteration loop. Furthermore, instead of aiming for complete autonomy, we advocate for flexible levels of shared autonomy, enabling human-AI collaboration to be tailored dynamically for each phase of the cycle. The color-coded segments in Figure~\ref{fig:intro_ai_research} correspond to the phases of an example BCI project reported here, with the exception of automatic paper write-up, which was not addressed in this work.

In the following sections, we will structure the the main part of the manuscript along the schematic depicted in Figure~\ref{fig:intro_ai_research}, detailing each phase of the research process and how human-AI collaboration is implemented to maximize efficiency, adaptability, and the transfer of expertise.

\subsection{Idea generation}

The research ideas presented in Table~\ref{tab:research_questions} were generated using GPT-4o, as our underlying LLM. The model was provided with a detailed prompt to explore advancements in EEG-based motor imagery classification within the context of Brain-Computer Interfaces (BCIs) using the BCI Competition IV 2a dataset \cite{bci_competition_iv_2a}. We decided to use this specific dataset because it has been widely used, maybe the closest to a \textit{"MNIST of EEG"}, providing a rich background of published results. 

The prompt included instructions to identify key challenges, research gaps, motivations, and potential approaches across a range of topics, including domain adaptation, lightweight architectures, temporal dynamics, and explainability. GPT-4o synthesized the list of ideas by leveraging its training on a wide range of scientific and technical texts; notably here we did not fine-tune on more specific literature. Overall, this demonstrates the utility of LLMs in assisting researchers by rapidly generating diverse, structured, and actionable ideas that may have merit in guiding future investigations. Notably, the generated research ideas included indeed notoriously under-researched yet important topics such as the role of data preprocessing (c.f. example 10 in ~\ref{tab:research_questions}). On the other hand, an expert in the field in a single glance can find that all generated research questions have already been addressed by the research community (\cite{peterson_penalized_2019, meng_adversarial_2023, li_adversarial_2022, jin_novel_2022, kostas_bendr_2021, kasahara_neuroanatomical_2015, echtioui_novel_2021, saha_intra-_2020, ju_federated_2020, feng_decoding_2020, kessler_how_2024, duan_zero-shot_2020, ferrez_simultaneous_2008}). Therefore, the introduction of iterative refinement in the process of idea generation (e.g. like chain of thoughts \cite{lu2024ai_scientist}) and tools for novelty verification (through Semantic Scholar API) and quantification are necessary. Clearly, the scientific ideas generation stage itself could also benefit from more sophisticated approaches. For instance, leveraging creative templates grounded in evolutionary computation and open-endedness principles offers significant potential. These approaches, as exemplified in works by \cite{lu2024ai_scientist} and \cite{faldor2024omniepicopenendednessmodelshuman}, use iterative, mutation-driven frameworks to refine and evolve novel ideas, mimicking the exploratory processes of natural evolution. Furthermore, combining knowledge graphs with LLM, as demonstrated by \cite{gu2024interestingscientificideageneration}, allows for the synthesis of diverse interdisciplinary ideas. Another promising approach is curiosity-driven learning, as proposed by \cite{pathak2017curiositydrivenexplorationselfsupervisedprediction}, which emphasizes exploring the unknown and prioritizing novelty over immediate utility.

However, it was also observable that many of the proposed topics were under-specified and lacked exact knowledge of what has already been done and reported in the relevant literature. For example, the suggestion was to explore lightweight neural networks that achieve state-of-the-art accuracy while being optimized for low-power devices, which is a promising topic, but one that is already well-trodden in the field. Studies like the widely known EEGNet paper  \cite{lawhern2018eegnet} have specifically addressed this challenge by designing compact, efficient neural networks tailored to EEG data. Therefore, while this idea aligns with key goals in the field, it lacks specificity in identifying novel directions. For instance, future advancements might focus on extending lightweight models to handle more diverse datasets, integrating adaptive techniques that dynamically scale complexity based on resource availability, or innovating in model interpretability for constrained devices. Without domain-specific fine-tuning, the model had also difficulties in retrieving the published SoA in decoding accuracies on the BCI Competition IV 2a dataset. While a re-implementation of the FBCSP winning model of the original BCI Competition IV yielded 67.8\% accuracy across the 4 motor imagery classes, and convnets allowed for a gain of approx. 4\% \cite{schirrmeister2017deep}. In the meanwhile, accuracies close to 100\% have been reported (e.g., 97.61\% in \cite{xie2023classification}). 

On this background, we decided for the following, human-specified topic for the first ChatBCI project: \textbf{What do the close to 100\% decoding accuracies reported on the BCI Competition IV 2a dataset mean?} To which extent do they reflect sophistication of the brain signal decoding methodology, or possibly also over-fitting to this (relatively small) dataset, and/or contributions from non-brain signals (artifacts which may contain class-specific information and might be easier to decode than brain signals)? The BCI Competition IV 2a data set is particularly interesting in this context, as it was part of the original competition instructions that since three EOG channels were provided, it was \textit{"required to remove EOG
artifacts before the subsequent data processing using artifact removal techniques such as highpass filtering or linear regression"} \cite{bci_competition_iv_2a}. The winning entry of the competition and based on their approach, also subsequent works on the same data set used a 4-Hz high-pass filter to address this requirement \cite{ang2012filter} \cite{tangermann2012review} \cite{schirrmeister2017deep}. 

Thus, for our first ChatBCI project, we set two objectives:  
\begin{itemize}
    \item Perform an exploratory data analysis of the BCI Competition IV 2a data set
    \item Design, implement and test a convolutional neural network (CNN) decoder with a good tradeoff of simplicity and hence interpretability on the one side, and meaningful decoding results in terms of decoding accuracies on the other, including a corresponding within-subject training loop.  
\end{itemize}

We addressed these objectives using the ChatBCI toolbox in a human-AI collaborative framework (Stage 2 \textit{"Experiment Iteration"} according to our generic research cycle plan ~\ref{fig:intro_ai_research}). 

\begin{table}
    \caption{Generated research questions, gaps, motivations, and approaches for advancing EEG-based motor imagery classification (10 consecutively generated examples).}
    \centering
    \begin{tabular}{p{0.5cm}p{3.5cm}p{3.5cm}p{3.5cm}p{3.5cm}}
        \toprule
        \textbf{ID} & \textbf{Research Question} & \textbf{Gap} & \textbf{Motivation} & \textbf{Approach} \\
        \midrule
        1 & What are the optimal EEG frequency bands for decoding, and how do they vary across subjects? & Inconsistent findings on band contributions. & Personalization can improve performance. & Perform detailed frequency band analysis. \\
        \midrule
        2 & How can adversarial robustness techniques improve the reliability of EEG classifiers? & Limited research on adversarial robustness in BCIs. & Reliable BCIs are essential for sensitive domains. & Simulate adversarial attacks and develop defenses. \\
        \midrule
        3 & Can graph-based representations of EEG signals improve motor imagery classification? & EEG relationships are underutilized in flat feature models. & Graphs can capture spatial-temporal relationships. & Use GNNs and evaluate performance. \\
        \midrule
        4 & How can self-supervised learning reduce the need for labeled data in EEG decoding? & Labeled data is scarce and expensive. & Self-supervised methods can leverage large unlabeled datasets. & Apply techniques like contrastive learning and test results. \\
        \midrule
        5 & What role does individual variability in brain anatomy play in decoding motor imagery? & EEG decoding often assumes uniformity across individuals. & Personalized BCIs can improve accuracy. & Analyze variability and develop normalization strategies. \\
        \midrule
        6 & How can dynamic ensemble methods improve the robustness and accuracy of motor imagery classification? & Static ensembles do not adapt to data properties. & Dynamic ensembles can tailor predictions to the data. & Develop adaptive ensemble methods and test performance. \\
        \midrule
        7 & What is the impact of session-to-session variability, and how can it be mitigated? & Session variability affects model performance. & Reliable BCIs need to function consistently over time. & Apply adaptation techniques and evaluate performance improvements. \\
        \midrule
        8 & Can federated learning improve decoding while preserving privacy? & Most approaches require centralized data. & Privacy-preserving training can enable collaborative BCIs. & Implement federated learning frameworks and test cross-subject models. \\
        \midrule
        9 & How can cross-frequency coupling (CFC) features improve decoding accuracy? & CFC is underexplored in motor imagery. & CFC can reveal richer brain dynamics. & Extract CFC features and integrate into models. \\
        \midrule
        10 & What are the effects of different EEG preprocessing pipelines on decoding performance? & No consensus on the best preprocessing pipeline. & Standardization can improve reproducibility. & Compare pipelines and evaluate their effects. \\
        \midrule
        11 & Can few-shot learning enable accurate decoding with minimal training data? & Most models require substantial data, impractical for new users. & Few-shot learning reduces the burden of data collection. & Implement few-shot methods like prototypical networks. \\
        \midrule
        12 & How can real-time feedback loops improve motor imagery classification during online experiments? & Feedback is minimally studied in offline datasets. & Real-time feedback could help users refine mental strategies. & Simulate real-time feedback and test its impact on decoding. \\
        \bottomrule
    \end{tabular}
    \label{tab:research_questions}
\end{table}

\subsection{Experimental interation}

Basic steps such as such data epoching, re-referencing, or filtering could be conveniently implemented, typically with a few prompts per step. The key functionality of the ChatBCI toolbox enabling this smooth workflow is the knowledge base with the relevant background knowledge and specific instructions, provided at different levels of detail and aggregation, such that the relevant information can be efficiently kept in the LLM context. 

More fine-grained instructions were (as expected) needed for data visualization. Figures ~\ref{fig:erp_figure}, ~\ref{fig:erp_figure_zoom} (A and B) and ~\ref{fig:decoding_figure} are collaboratively ChatBCI-generated. Code generation and running the experiments were the two steps with the highest level of AI autonomy (see Fig. ~\ref{fig:intro_ai_research}) 

\subsubsection{Exploratory data analysis and data validation}

Analysis for data exploration and validation let to several interesting observations. ChatBCI suggested initially three starting steps for data validation and exploration. \textit{First}, examine the basic signal statistics to understand the distribution and variability of the EEG data across classes. This involves calculating the mean, standard deviation, and variance for each channel and class, identifying systematic differences in signal variance, and detecting outliers, which may indicate artefacts. \textit{Second}, analyze event-related potentials (ERPs) to investigate time-locked neural activity by segmenting the data around key events, calculating average ERP waveforms for each class, and identifying distinctive features such as P300 peaks while ensuring the observed differences are neural rather than artefact-related. \textit{Third}, explore class-specific differences in the frequency domain by computing power spectral density (PSD) for each class and channel, identifying frequency bands where power varies, and detecting potential artefacts such as line noise or muscle activity. 

\begin{figure}[h!]
    \centering
    \includegraphics[width=0.8\textwidth]{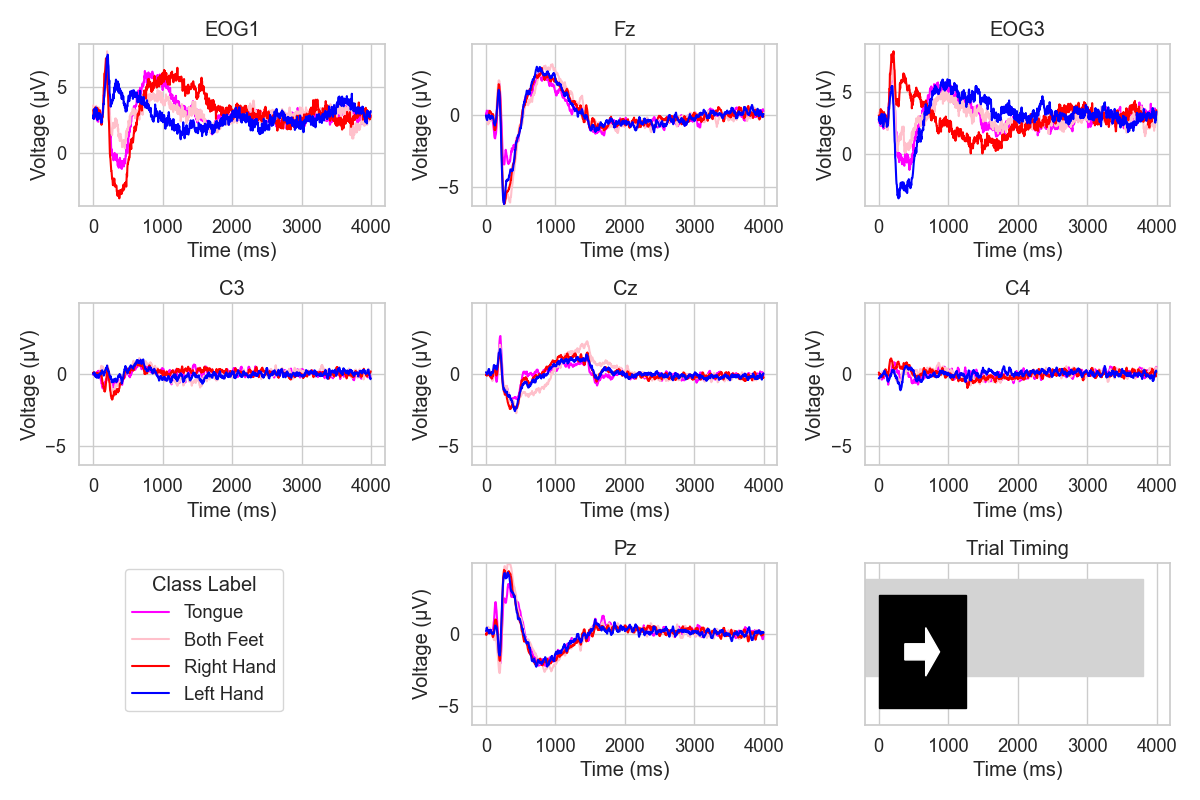}
    \caption{ERP waveforms across all trials of all subjects' training data. Trial timing: The cue in the form of an arrow pointing either to the left, right, down or up, corresponding to one of the four classes left hand, right hand, foot or tongue) appeared and stayed on the screen for the duration indicated by the black box (1.25 s). Grey box: Time window of fixation cross presentation.}
    \label{fig:erp_figure}
\end{figure}

The questions whether we can indeed expect a P300 as the dominating effect in the ERPs based on this dataset, and what we can expect, revealed a mixture of accurate and wrong assumptions (e.g., correctly expecting visually and auditory-evoked potentials (VEPs, AEPs) elecited by the respective cues, but also expecting a readiness potential which in fact is associated with self-paced tasks \cite{ball1999higher_order_motor}), illustrating the necessity of expert knowledge transfer. 

ERP results (EEG common average rereferenced, 40 Hz low-pass filtered) are shown in Fig. ~\ref{fig:erp_figure}, for 5 EEG and 2 EOG channels. 

Next, we asked for an interpretation of this plot with the prompt \textit{"this figure shows plots of the ERPs (mean across all trials of all subjects for each of the 4 classes). What do we see here, how can we interpret this result?"} with the response which we fully reproduce here (in blue font): 

\vspace{0.5cm} 
{\color{blue!80!black}
"The provided figure shows \textbf{event-related potentials (ERPs)} averaged across all trials and subjects for each of the 4 motor imagery classes (tongue, both feet, right hand, and left hand). Each subplot corresponds to a different electrode or channel (e.g., EOG1, Fz, C3, Cz, C4, Pz), with voltage plotted as a function of time (0--4000 ms).

\section*{Key Observations}

\textbf{1. Cue-Evoked Potentials:}
\begin{itemize}
    \item For channels like \textbf{Fz, Cz, and Pz}, there is a clear deflection around \textbf{200--300 ms} after the cue onset, which could represent the early processing of the cue. This is consistent with components such as \textbf{N1/P2} (cue-related processing).
    \item These components are visible across all classes, indicating that these potentials are \textbf{not class-specific} but instead reflect the shared cognitive processing of the presented cue.
\end{itemize}

\textbf{2. Motor-Related Activity:}
\begin{itemize}
    \item Channels \textbf{C3, Cz, and C4} are primarily associated with motor areas:
    \begin{itemize}
        \item For \textbf{C3} (left motor cortex), the \textbf{right hand} class shows a small deviation compared to other classes, consistent with contralateral activity.
        \item For \textbf{C4} (right motor cortex), the \textbf{left hand} class exhibits a subtle difference, though less pronounced.
    \end{itemize}
    \item Both feet and tongue motor imagery do not show distinct deflections in these motor channels, likely due to their bilateral or non-cortical representation.
\end{itemize}

\textbf{3. Eye Movement Artefacts:}
\begin{itemize}
    \item \textbf{EOG1 and EOG3} channels clearly show larger deviations, especially for the \textbf{tongue} and \textbf{both feet} classes. This suggests that these classes may involve stronger or unintended \textbf{eye movement artefacts}, particularly in the early time window (0--1000 ms).
    \item These artefacts could partially overlap with class-specific ERP components and may need to be addressed (e.g., through filtering or artifact rejection).
\end{itemize}

\textbf{4. Overall Class Differences:}
\begin{itemize}
    \item While class-specific differences are subtle, the \textbf{right hand} and \textbf{left hand} classes show slight distinctions in the \textbf{C3 and C4} channels, corresponding to their expected motor areas.
    \item The \textbf{both feet} and \textbf{tongue} classes show fewer clear class-specific ERP features, and their differences may instead emerge more clearly in the frequency domain (e.g., mu/beta desynchronization).
\end{itemize}

\section*{Interpretation}

\begin{itemize}
    \item \textbf{Cue-Related Potentials:} These are robust across all classes, primarily seen in midline channels (e.g., Fz, Cz, Pz), reflecting the shared processing of visual stimuli.
    \item \textbf{Motor-Specific Activity:} Subtle differences in the motor-related channels (C3, C4) align with the expected contralateral activation for hand motor imagery tasks.
    \item \textbf{Artefact Contribution:} Large deviations in EOG channels suggest that some ERPs might be influenced by eye movement artefacts, particularly for classes like tongue and both feet."
\end{itemize}
}

\begin{figure}[h!]
    \centering
    \includegraphics[width=0.8\textwidth]{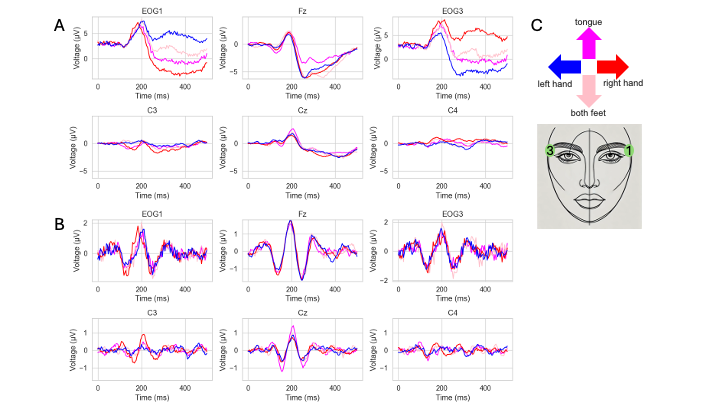}
    \caption{(A) Zoom-in from Fig. ~\ref{fig:erp_figure}. (B) Same with 4-Hz high-pass filter. (C) Direction of the arrows used as cues for the 4 classes and colored corresponding to (A) and (B); placement of the EOG1 and EOG3 channels as described in \cite{bci_competition_iv_2a}.}
    \label{fig:erp_figure_zoom}
\end{figure}

This interpretation addresses many of the relevant aspects and in our opinion demonstrates a very promising starting point for a useful visual understanding of results in the EEG and BCI research context of. The three major signal sources that need to be considered (cue-evoked, movement-related, and ocular artifacts) are correctly identified and discussed, with several valid points such as that class differences can be expected to be clearer in the frequency domain. However, the conclusion that ocular effects mainly show in the tongue and feet movement classes is not likely given the observations that can here be made. The direction of the arrows used as cues \footnote{Placement and coloring of the arrows in ~\ref{fig:erp_figure_zoom} serves for illustration, not implying a similar placement or coloring in the original study, where arrows were, based on the available information, all paced centrally and in the same color, but pointing to different directions as show \cite{bci_competition_iv_2a}} together with the placement of the EOG channels (see Fig, ~\ref{fig:erp_figure_zoom} (C)) suggest that subjects reacted to visual cue presentation with a blink-saccade towards the tip of the different presented arrows; see also Fig. ~\ref{fig:erp_figure_zoom} (A). The fast transients at the onset of the presumed blink-saccade movement survived a 4-Hz high-pass filter (Fig. ~\ref{fig:erp_figure_zoom} (B)).  

In summary, already in its current early stage, ChatBCI proved highly useful for a practical example of a EEG data exploration, allowing for a substantially accelerated workflow. Analyzing a widely used motor imagery data set turned out to be a rich "learning opportunity" for ChatBCI (exemplarily showing that cues may induce class-specific effects, how to detect differential eye movements, etc.). At the current stage, our solution to make such learning experiences persistent is to include corresponding text documents in the ChatBCI knowledge base. In future versions, this approach may be supplemented by fine-tuning the underlying language models, and by equipping models with memory \cite{wolters2024memory} \cite{packer2024memgptllmsoperatingsystems}. 

One next step from here would be to check the published BCI literature to which extend our observations have already been reported, and to this end, implementing abilities for automated literature search and aggregation in ChatBCI. The results so far also support our initial choice of research questions and objectives (see Section 3.1). In the following, we briefly describe the first step with respect to Objective 2, design, implement and test a CNN EEG decoder. 

\subsubsection{LLM-generated deep neural network decoder and training loop}

\begin{figure}[h!]
    \centering
    \includegraphics[width=0.8\textwidth]{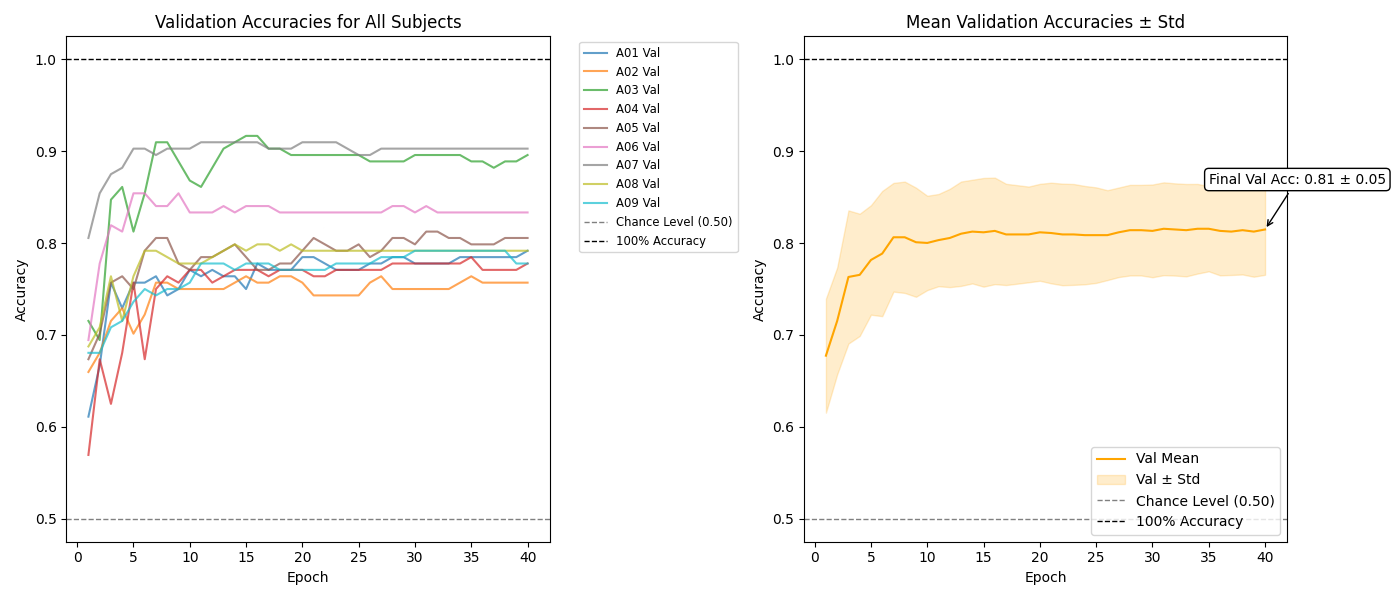}
    \caption{(A) Zoom-in from Fig. ~\ref{fig:erp_figure}. (B) Same with 4-Hz high-pass filter. (C) Direction of the arrows used as cues for the 4 classes and colored corresponding to (A) and (B); placement of the EOG1 and EOG3 channels.}
    \label{fig:decoding_figure}
\end{figure}

Not surprisingly, given the well-documented abilities of LLMs in code generation (\cite{rozière2024codellamaopenfoundation}), ChatBCI was able to design and implement both deep networks for EEG decoding and working training loops, including plausibly motivated data-augmentation strategies. Figure ~\ref{fig:decoding_figure} shows results obtained with a simple network consisting of a sequence of spatial and temporal convolutions, using batch normalization, dropout, and a SWISH activation function, with within-subject training using the original competition data split. 

Results showed clear learning on the validation set across all subjects, reaching accuracies, on the one hand, clearly above chance, and on the other hand, also substantially below previous reports \cite{zhang_eeg-inception_2021}. We consider our simple models as a good starting point for, in a next step, implementing interpretability tools, and gradually moving to more complex decoding models. Setting up the network as well as the training loop used in this example, in total required less than 10 prompts. 

Another next logical step would be the implementation of hyperparameter optimization and neural architecture search functionality in ChatBCI, for advancing to models with SoA accuracy or beyond. Note that this would potentially enable novel approaches for AutoML leveraging LLMs, without the necessity for expert knowledge of underlying AutoML libraries and toolboxes. We foresee that ChatBCI would also be particularly useful in BCI decoding benchmarking tasks, which require structured large-scale approaches which may especially align with the specific strengths of artificial compared to human intelligence. 

\section{Conclusions}

This paper introduced the ChatBCI toolbox as a practical and innovative solution to foster human-AI collaboration in the fields of Brain-Computer Interface (BCI) research and brain signal analysis. By adopting our \textbf{Janusian Design Principles}, we demonstrated how human-AI teaming can address challenges in BCI development, such as small datasets, complex signal dynamics, and knowledge transfer. Specifically:

\begin{enumerate}
    \item \textbf{Collaborative Human-AI Workspaces:}  
    The ChatBCI framework was built to integrate the strengths of both human expertise and AI capabilities, providing an adaptable, interactive workspace that supports all phases of scientific research, from idea generation to experimental analysis and result interpretation.

    \item \textbf{Application to Motor Imagery Decoding:}  
    Using the ChatBCI toolbox, we explored the BCI Competition IV 2a dataset and implemented a deep learning decoder in a collaborative, time-efficient manner. This process highlighted the utility of ChatBCI for accelerating workflows, identifying domain-specific challenges, and fostering human-AI co-learning.

    \item \textbf{Transfer of Expertise and Knowledge:}  
    The structured interaction framework enabled effective transfer of non-obvious, domain-specific knowledge from human experts to the AI, ensuring meaningful co-learning. The AI, in turn, proved to be a valuable partner, suggesting analyses, generating code, and adapting to user inputs.

    \item \textbf{Scalable Design for Broader Neurotechnological Applications:}  
    While demonstrated for a specific EEG dataset, the principles and methods behind ChatBCI are extensible to a wide range of neuroscience and neurotechnology challenges, laying a foundation for future advancements in brain research and AI-driven discovery.
\end{enumerate}

By merging the adaptability of AI systems with the nuanced understanding of human researchers, ChatBCI represents a step forward in achieving efficient and collaborative research processes. Our findings indicate that human-AI teaming, as enabled by ChatBCI, holds significant promise for accelerating scientific discovery while maintaining the interpretability and trust essential for impactful advancements in neuroscience and BCI.

We envision a future where in some scenarios brain-computer interfaces (BCIs) transcend traditional keyboard inputs, replacing them with direct interaction through physiological signals—a transformative goal of BCI research and development. For instance, advances like Meta's silent speech decoding demonstrate the potential of such technologies to enable seamless communication between humans and machines \cite{D_fossez_2023}. This innovation aligns perfectly with the evolving paradigm of human-AI collaboration, where neural interfacing could ensure seamless interaction and integration of human-AI teams. By integrating robust privacy and security measures (\citet{kapitonova2022frameworkpreservingprivacycybersecurity}), such systems could enable richer societal interactions and more immersive experiences in scientific discovery, healthcare, and beyond. The convergence of neurotechnology and AI promises not only to amplify human cognitive capabilities but also to unlock new paradigms for how humans and machines collaborate to solve complex problems.

Through the type of collaborative teamwork on brain data enabled by ChatBCI, AI systems may progressively develop a deep and holistic understanding of human brain function. By engaging with the complexities of brain signals, their multiscale dynamics, and their connections to cognition, behavior, and emotion, such AI systems may transcend conventional analytical approaches. This emerging paradigm has potential not only in neuroscience and neurotechnology but also fields such as medicine, psychology, and education. Furthermore, a \textit{"brain-grokking AI"} could unlock significant performance advantages in areas that rely on human brain function, including optimized mental health interventions and tailored cognitive training, underscoring the importance of human-AI collaboration as a cornerstone for advancing knowledge and unlocking new possibilities in science and technology.

\bibliographystyle{plainnat}
\bibliography{template}  

\end{document}